\documentclass[prc,twocolumn,epsfig,nofootinbib,floatfix]{revtex4}
\usepackage{graphics}
\usepackage{epsfig}
\usepackage{amsfonts}
\usepackage{amsmath}
\usepackage{bm}% bold math

\usepackage{color}

\begin{document}
\title{On the toroidal nature of the low-energy E1 mode}
\author{A. Repko$^1$, P.-G. Reinhard$^2$, V.O. Nesterenko$^3$, and J. Kvasil$^1$}

\affiliation{$^1$ Institute
of Particle and Nuclear Physics, Charles University, CZ-18000,
Praha 8, Czech Republic}
\affiliation{$^2$ Institut f\"ur Theoretische Physik II,
Universit\"at Erlangen, D-91058, Erlangen, Germany}
%\email{Paul-Gerhard.Reinhard@physik.uni-erlangen.de}
\affiliation{$^3$ Laboratory of Theoretical Physics, Joint
Institute for Nuclear Research, Dubna, Moscow region, 141980,
Russia} \email{nester@theor.jinr.ru}

\date{\today}

\begin{abstract}
The nature of E1 low-energy strength (LES), often denoted as a ``pygmy
dipole resonance'', is analyzed within the random-phase-approximation
(RPA) in $^{208}$Pb using Skyrme forces in fully self-consistent
manner.  A first overview is given by the strength functions for the
dipole, compressional and toroidal operators. More detailed insight is
gained by averaged transition densities and currents where the latter
provide very illustrative flow pattern. The analysis reveals clear
isoscalar toroidal flow in the low-energy  bin 6.0-8.8 MeV of the LES
and a mixed isoscalar/isovector toroidal/compression flow in the
higher bin 8.8-10.5 MeV. Thus the modes covered by LES embrace both
vortical and irrotational motion.  The simple collective picture of
the LES as a ``pygmy'' mode (oscillations of the neutron excess
against the nuclear core) is not confirmed.
\end{abstract}


\maketitle

During the last decade we observe an increasing interest in
low-energy E1 strength (LES), for a recent review see
\cite{Paar_07}. This interest is caused by a possible relation of
LES to the neutron skin in nuclei and density dependence of the
nuclear symmetry energy. This in turn may be important for
building the isospin-dependent part of the nuclear equation of
state and various astrophysical applications \cite{Carbone_10}.
Several different views of the LES origin come together. Most
often the LES is interpreted as a "pygmy dipole resonance" (PDR)
modeled as the oscillation of the neutron excess against the
nuclear core \cite{Is92,Vr01,Paar_07,Maza12}. There are, however,
serious objections against such a simplistic collective picture
\cite{Yu12,Rei12}. In fact, the landscape of LES may be much
richer. It can embrace also the toroidal resonance (TR)
\cite{Du75,Se81} and anisotropic compression resonance (CR)
\cite{Ha77,St82} which both are of actual interest \cite{Paar_07}.
After exclusion of nuclear center-of-mass (c.m.) motion, the TR
and CR dominate in the isoscalar (T=0) channel and constitute the
low- and high-energy branches of the isoscalar giant dipole
resonance (ISGDR). Following recent microscopic studies
\cite{Kv11,Kv12}, the TR dominates in the LES region and the CR,
being strongly coupled to TR, also significantly contributes
there.
%%%%%%%%%%%%%%
% Figure 1
%%%%%%%%%%%%
\begin{figure}
\includegraphics[width=\linewidth]{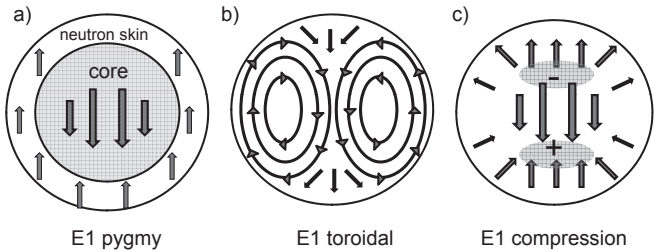}
\caption{Schematic velocity fields for the E1 pygmy (a), toroidal (b),
and high-energy compressional (c) flows. The driving field is directed along z-axis.
The arrows indicate only directions of the flows but not their strength. In the plot (c),
the compression (+) and decompression (-) regions, characterized by increased and decreased
density, are marked.}
\end{figure}
The basic flow patterns of these three modes are shown schematically in
Fig. 1. The panels illustrate the PDR oscillations of the neutron skin
against the core (a), the typical vortices of the TR (b), and
the dipole-compressional pattern of the CR (c). The latter can be viewed
as oscillation of surface against core and thus shares some similarity
with the PDR picture of panel (a).  Unlike the irrotational PDR and CR,
the TR is purely vortical in the hydrodynamical (HD) sense
\cite{Kv11,Kv12}. Thus we see that the LES can involve quite
different flows, vortical and irrotational.

Despite a great number of publications on PDR, TR and CR
\cite{Paar_07}, their possible interplay in LES was only
occasionally discussed. There is a study within the
quasiparticle-phonon model which discusses PDR and TR side by side
\cite{Ry02}. It was shown that the vortical strength \cite{Ra87}
is peaked in the LES region and the isovector LES velocity field
is mainly toroidal.  Nevertheless, because of the dominant
contribution of the neutron skin to the surface transition density
and thus to B(E1,T=1), the familiar PDR picture of LES was
maintained \cite{Ry02}. Similar arguments in favor of the PDR treatment were
earlier presented in relativistic \cite{Vr01} and Skyrme
\cite{Maza12} mean-field calculations, and taken up in most of
subsequent publications.  Recent explorations question the simple
PDR-type collectivity of LES \cite{Kv12,Yu12,Rei12}, though without
analysis of LES velocity fields.

It is the aim of this paper to give a more thorough exploration of
the interplay and structure of low lying dipole modes. Not only
strength functions and transition densities
but also the mode flow patterns will be considered. As we
will see, the actual LES flow is predominantly of mixed TR/CR
character. This conclusion may have far-reaching consequences for
the information content of LES. If vorticity dominates, then
only a minor irrotational fraction of LES is relevant for the
nuclear symmetry energy and related problems (as was also worked
out recently using the correlation analysis \cite{Rei12}).

Our study is performed for $^{208}$Pb using the Skyrme RPA
approach with the techniques of \cite{Rei92}. The method is fully
self-consistent as both the mean field and residual interaction
are derived from the Skyrme functional \cite{Skyrme,Vau72,Ben03}.
The RPA residual interaction takes into account all the terms of
the Skyrme functional including the Coulomb (direct and exchange)
energy. The center-of-mass correction (c.m.c.)  is implemented for
the ground state and  T=0 dipole excitations. The parameterization
SLy6 \cite{Sly6} is used which provides a satisfactory description
of E1(T=1) strength in heavy nuclei \cite{nest_PRC_08}. The
calculations are done in a 1D spherical coordinate-space grid with
mesh size 0.3 fm and a calculation box of 21 fm. A large
configuration space including $1ph$ states up to $\sim 35$ MeV and
additional fluid dynamical basis modes is used. The later allows
to include global polarization effects up to $\sim$200 MeV
\cite{Rei92}, correctly extract the c.m. motion, and fully exhaust
the Thomas-Reiche-Kuhn sum rule.

The excitation modes are first characterized by their strength
function
\begin{equation}
\label{eq:strength_function}
  S_{\alpha}(E1; \omega) = 3
\sum_{\nu} \omega_{\nu}^l
  |\langle\Psi_\nu|\hat{M}_{\alpha}(E10)|\Psi_0\rangle|^2
  \zeta(\omega,\omega_{\nu})
\end{equation}
where $ \zeta(\omega,\omega_{\nu}) = \Delta(\omega_\nu)/[2\pi
  [(\omega- \omega_{\nu})^2+\Delta(\omega_\nu)^2/4]] $ is a Lorentzian
weight with energy-dependent smoothing width $\Delta
(\omega_{\nu})= \text{max}\{ 0.4 \; \text{MeV}, (\omega_{\nu}-8 \;
\text{MeV})/3)\}$, for details see \cite{Kva_IJMPE_11}. Further,
$\Psi_0$ is the RPA ground state (g.s.) while $\nu$ runs over the
RPA spectrum with eigen-frequencies $\omega_{\nu}$ and
eigen-states $|\Psi_\nu\rangle$. The $\hat{M}_{\alpha}(E1\mu)$ is
the transition operator of the type $\alpha = \{\text{E1, tor,
com}\}$.

For E1(T=1) transitions ($\alpha\equiv$E1), we consider the
ordinary E1 operator ($\propto rY_{1\mu}$) with effective charges
$e_{\mathrm{1}}^p=N/A$ and $e_{\mathrm{1}}^n=-Z/A$ and the
strength (\ref{eq:strength_function}) is weighted by the energy,
i.e. $l=1$. For $\alpha $= tor,com, we implement
$e_{\mathrm{0}}^p=e_{\mathrm{0}}^n=1$ for T=0
($e_{\mathrm{1}}^p=-e_{\mathrm{1}}^n=1$ for  T=1) and no energy
weight ($l=0$).

The TR  and CR  operators used in
(\ref{eq:strength_function}) read \cite{Kv11}
\begin{eqnarray}
\label{TM_curl} && \hat M_{\text{tor}}(E1\mu) =
\\
\nonumber
&&
-\frac{1}{10\sqrt{2}c} \int d^3r
[r^3-\frac{5}{3}r \langle r^2\rangle_0]
\vec Y_{11\mu}(\hat{\vec r})
\cdot (\vec{\nabla}\!\times\!\hat{\vec
j}_{c}(\vec r)) \; ,
\\
\label{CM_div}
&& \hat M_{\text{com}}(E1\mu) =
\\ \nonumber
&&
 -\frac{i}{10c} \int d^3r
[r^3\!-\!\frac{5}{3}r \langle r^2\rangle_0] \;
Y_{1\mu}(\hat{\vec r})\; (\vec{\nabla} \cdot \hat{\vec j}_{c}(\vec r)) \; ,
\end{eqnarray}
where $\hat{\vec j}_{c}(\vec r))$ is the operator of the
convection nuclear current, $\vec{Y}_{11\mu}(\hat{\vec r})$ and
$Y_{1\mu}(\hat{\vec r})$ are vector and ordinary spherical
harmonics.  The terms with the g.s. squared radius $\langle
r^2\rangle_0 = \int\:d^3r \:r^2 \rho_0(\vec{r})/A$ account for the
c.m.c., $\rho_0(\vec{r})$ is the g.s. density.  Note that we
describe CR and TR operators on the same footing using the current
operator. There is the direct relation \cite{Kv11} $
\hat{M}_{\text{com}}(E1\mu)=-\hat{M}'_{\text{com}}(E1\mu)\omega/c$
between the CR current-dependent operator (\ref{CM_div}) and its
familiar density-dependent counterpart
\begin{equation}
\label{CMp_oper}
 \hat M'_{\text{com}}(E1\mu) =  \frac{1}{10} \int d^3r
\hat{\rho}(\vec r)
[r^3-\frac{5}{3}\langle r^2\rangle_0 r] Y_{1\mu}(\hat{\vec r}) \; ,
\end{equation}
where $\hat{\rho}(\vec r)$ is the density operator.

The operators (\ref{TM_curl})-(\ref{CMp_oper}) are derived as
second-order $\sim r^3 Y_{1\mu}$ terms in the low-momentum
expansion of the ordinary E1 transition operator
\cite{Du75,Se81,Kv11}.  Despite its second-order origin, TR and CR
dominate in the E1(T=0) channel where the leading c.m. motion
driven by the operator $r Y_{1\mu}$ is removed as being the
spurious mode.  Following (\ref{TM_curl})-(\ref{CM_div}), TR and
CR deliver information on the curl $\vec{\nabla} \times \hat{\vec
j}_{c}$ and divergence $\vec{\nabla} \cdot \hat{\vec j}_{c}$ of
the nuclear current.  As shown in \cite{Kv11}, the corresponding
velocity operators indicate that TR  is purely vortical and  CR is
irrotational.
%%%%%%%%%%%%%%
% Figure 2
%%%%%%%%%%%%
\begin{figure}
\includegraphics[width=6.3cm]{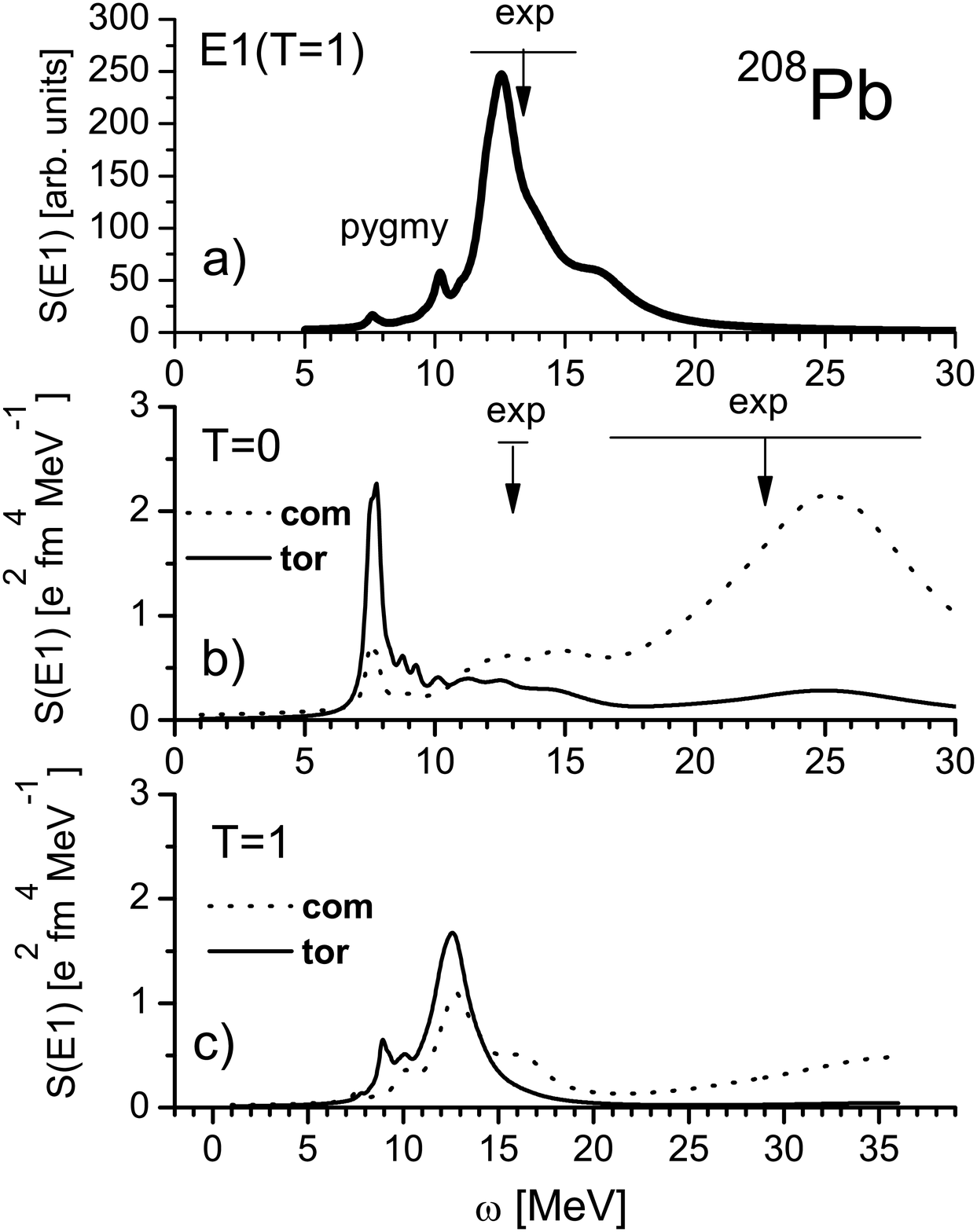}
%\includegraphics[width=\linewidth]{fig9.eps}
%\vspace{3mm}
\caption{Strength functions calculated within RPA
with the force SLy6. (a) E1(T=1) giant resonance  The line and
arrow indicate the experimental width and energy centroid of the
resonance \protect\cite{GDR_exp}. The pygmy region is marked.
(b) Toroidal
(solid line) and  compression (dotted line) E1(T=0) strengths. The
widths and energy centroids of the low- and high-energy branches
of ISGDR observed in ($\alpha, \alpha'$) reaction
\protect\cite{Uchida_PRC_04} are denoted.
c) The same as in the plot (b) but in T=1 channel.}
\end{figure}

The strength functions (\ref{eq:strength_function}) are shown in
Fig. 2.  In panel a), we see a good agreement of the computed
giant dipole resonance (GDR) with the experiment \cite{GDR_exp},
which confirms the accuracy of our description. For the LES region
6-10.5 MeV (marked as pygmy), we get two peaks at 7.5 and 10.3 MeV
in accordance with previous RMF calculations \cite{Vr01}. Panels
b) and c) show TR and CR strengths in T=0 and T=1 channels. For
T=0, the TR and CR  are believed to constitute the low- and
high-energy parts of the ISGDR \cite{Paar_07}. Our results
somewhat deviate from the experimental $(\alpha,\alpha')$ data
\cite{Uchida_PRC_04}, but a similar discrepancy takes place in
almost all theoretical studies \cite{Paar_07}. Perhaps, at 12-14
MeV not the TR but the low-energy CR bump (see panel b)) is
observed. The discrepancy for the high-energy CR may be caused by
its sensibility to the calculation scheme.  What is important for
our aims, the LES region should certainly host the dominant and
strongly peaked part of TR(T=0), the left flank of TR(T=1), and a
non-negligible low-energy fraction of CR. In other words, we
expect here a complicated interplay of several modes.

To understand the LES structure, we need more detailed observables
 than the strength distribution. In the following, we
will consider transition densities (TD) $\delta \rho_{\nu} (r) =
\langle\Psi_\nu|\hat{\rho}|\Psi_0\rangle$ and current transition
densities (CTD) $\delta \vec{j}_{\nu} (\vec{r}) =
\langle\Psi_\nu|\hat{\vec{j}}_c|\Psi_0\rangle$ (analogous to
velocity fields).  As we have in $^{208}$Pb a high density of
states, it is not worth to look at the pattern of individual
states $\nu$, which can vary from state to state and easily hide
common features of the flow. Thus we will consider transition
densities and velocity fields averaged over given energy
intervals. Incoherent averaging requires expressions which are
bi-linear in the excited states $|\Psi_\nu\rangle$.  This is
achieved by summing TD and CTD weighted by the matrix elements
$D_{T
  \nu}=\langle \nu|\hat{D}_T(E1)|0\rangle$ of a probe operator
$\hat{D}_T(E1)$:
\begin{equation}\label{sum_td}
 \delta \rho^{(D)}_{\beta}(\vec{r}) = \sum_{\nu \epsilon [\omega_1, \omega_2]}
  D_{T \nu}^*
  \sum_{q=n,p} e^{q}_{\beta} \delta \rho_{\nu}^{q}(\vec{r})
  \;,
\end{equation}
\begin{equation}
\label{sum_tcd}
  \delta \vec{j}^{(D)}_{\beta}(\vec{r}) = \sum_{\nu \epsilon [\omega_1, \omega_2]}
  D_{T \nu}^*
   \sum_{q=n,p} e^{q}_{\beta}\delta \vec{j}_{\nu}^{q}(\vec{r})
  \;.
\end{equation}
The sums in (\ref{sum_td})-(\ref{sum_tcd}) involve all the RPA states
$|\nu\rangle$ in the energy interval $[\omega_1, \omega_2]$.  Since
states $|\nu\rangle$ contribute twice (to $D_{T \nu}^*$ and transition
densities $\delta \rho_{\nu}^{q}$/ $\delta \vec{j}_{\nu}^{q}$), the
expression is independent of the phase of each state
$|\Psi_\nu\rangle$ as it should be.  In
(\ref{sum_td})-(\ref{sum_tcd}), the index $\beta$ = p,n,0,1 defines
the type of TD or CTD (neutron, proton, T=0, T=1) by the proper choice
of the effective charges: $e^{p}_{p}=1, e^{n}_{p}=0$; $e^{p}_{n}=0,
e^{n}_{n}=1$; $e^{p}_{0}=e^{n}_{0}=1$; $e^{p}_{1}=N/A,
e^{n}_{1}=-Z/A$. We use  two different dipole probe operators: the
isovector $\hat{D}_1=(N/A)\sum_i^Z (r Y_1)_i -(Z/A)\sum_i^N (r
Y_1)_i$ relevant for reactions with photons and electrons, and
isoscalar compressional $\hat{D}_0=\sum_i^A (r^3 Y_1)_i$ relevant for
$(\alpha, \alpha')$ reaction. Due to $ D_{T \nu}^*$ weights, the
contribution of RPA states with a large $D_T$ strength is enhanced.

%%%%%%%%%%%%%
% Figure 3
%%%%%%%%%%%%
\begin{figure}
\includegraphics[width=7.5cm]{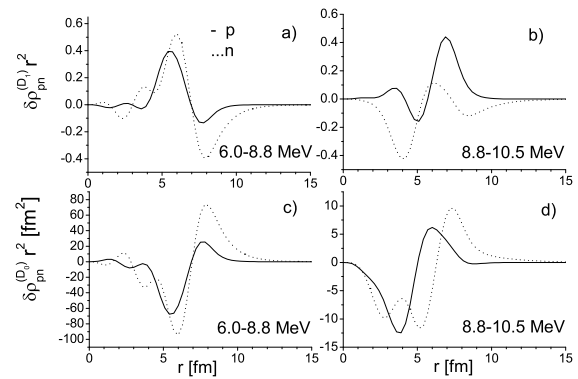}
\caption{ Summed $r^2$-weighted proton and neutron TD $\delta \rho^{(D_1)}_{p,n}$ (a-b)
and $\delta \rho^{(D_0)}_{p,n}$ (c-d) at the energy intervals 6.0-8.8 MeV (left)
and 8.8-10.5 MeV (right).}
\end{figure}
In Fig. 3, the TD summed over two bins of the LES, 6.0-8.8 MeV and
8.8-10.5 MeV, are shown. One sees that, up to a scale factor, the
TD for the probes $D_1$ and $D_0$ are rather similar, especially
at 6.0-8.8 MeV. Panels a) and c) show that at 6.0-8.8 MeV the
protons and neutrons oscillate in phase in the interior area 4 - 7
fm (isoscalar flow of the core) but neutrons dominate at larger
distances $r>7$ fm (contribution of the neutron excess). Due to
$r^2$-factor, mainly the surface area $r>7$ fm contributes to the
B(E1) $\propto \int dr r^2 \delta\rho(r)$. This would favor the
simple PDR picture of neutron-core oscillations.  At the higher
energy, 8.8-10.5 MeV, we see mainly isovector motion at 6-8 fm and
again dominance of neutrons at $r>8$ fm. So, LES here is more
isovector and does not support the PDR picture. Furthermore, the
isoscalar $D_0$ leads to much weaker TD at 8.8-10.5 MeV relative
to the region 6.0-8.8 MeV. This is probably caused by the fact
that LES at 6.0-8.8 MeV is mainly isoscalar thus gaining more from
the isoscalar weight. Note that the $r^2$-factor amplifies the
pattern in the nuclear surface and damps it in the interior area
at $r<$ 4 fm. The TD in Fig. 3 indicate that there are sizable
effects in the interior. This will be corroborated by the flow
pictures below.

A thorough analysis of excitations should also look at CTD which reveal
even more details than mere TD.
The CTD for dipole states $\lambda\mu=10$ are presented in
Figs. 4-7. In Fig. 4, the fields for the isovector GDR (10.5-15 MeV) and
high-energy isoscalar CR (22-30 MeV) are given as reference
examples. They show typical GDR and CR flows, see \cite{Ry02,Vr00,Misicu06}
for a comparison. In the CR case, the
compression and decompression zones along the $z$-axis are visible,
as in Fig. 1c). These plots for well known modes serve as benchmark
and assert the validity of our prescription.
%%%%%%%%%%%%%
% Figure 4
%%%%%%%%%%%%
\begin{figure}
\includegraphics[width=\linewidth]{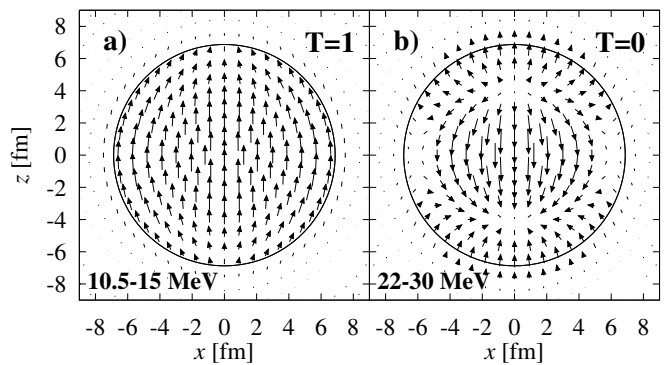}
\vspace{5mm}
\caption{Summed CTD (a) $\delta \vec{j}^{(D_1)}_{1}$ for GDR in the
bin 11-15 MeV and (b) $\delta \vec{j}^{(D_0)}_{0}$ for CR(T=0)
in the bin 22-30 MeV (b).}
\end{figure}
%%%%%%%%%%%%%
% Figure 5
%%%%%%%%%%%%
\begin{figure}
\includegraphics[width=8.3cm]{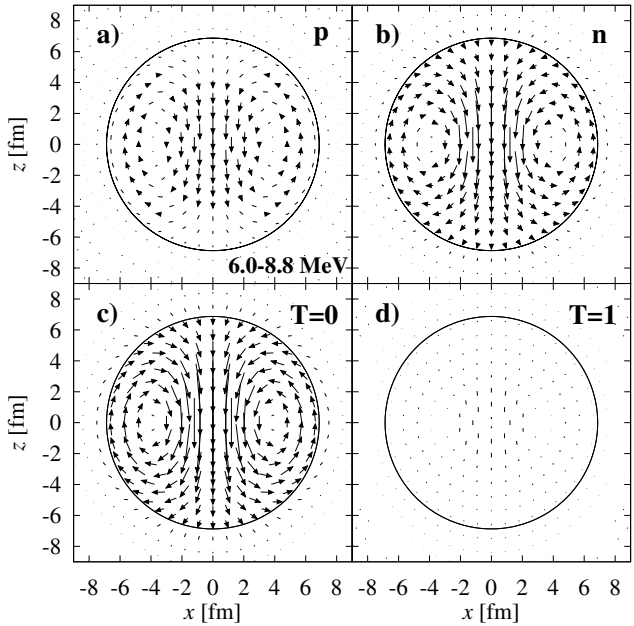}
\caption{Proton (a), neutron (b), T=0 (c) and T=1 (d)
summed CTD $\delta \vec{j}^{(D_1)}_{\beta}$ in the bin 6.0-8.8 MeV. }
\end{figure}
%%%%%%%%%%%%%
% Figure 6
%%%%%%%%%%%%
\begin{figure}
\includegraphics[width=8.3cm]{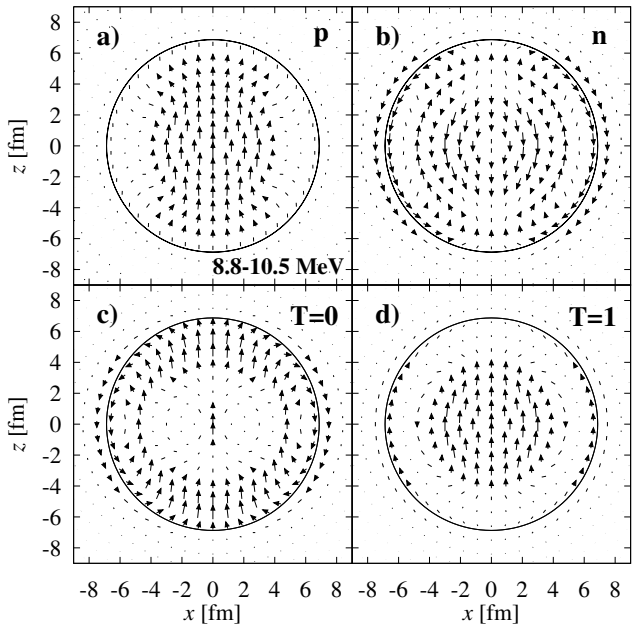}
\caption{The same as in Fig. 5 but for the bin
8.8-10.5 MeV.}
\end{figure}
%%%%%%%%%%%%%
% Figure 7
%%%%%%%%%%%%
\begin{figure}
\includegraphics[width=8.3cm]{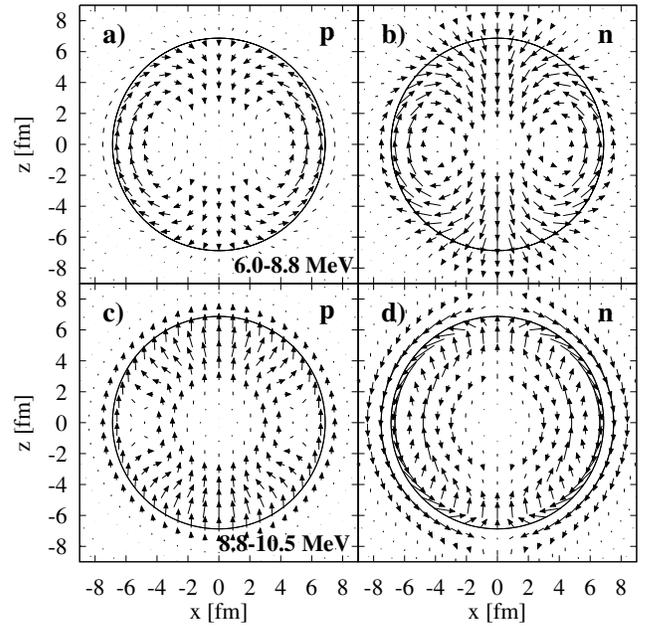}
\caption{The $r^2$-weighted proton and neutron  $\delta \vec{j}^{(D_1)}_{\beta}$
in the bins 6.0-8.8 MeV (a)-(b) and 8.8-10.5 MeV (c)-(d).
}
\end{figure}

In Figs. 5-7 the fields for the two LES bins, 6.0-8.8 and 8.8-10.5
MeV, are depicted.  Since $D_1$ and $D_0$ fields look, up to a
scale factor, rather similar (especially for the bin 6.0-8.8 MeV),
we will show further on only $D_1$ weighted CTD. In every figure,
the actual CTD scale (common for all the plots) is adjusted for
better view. Arbitrary units for CTD are used.

Fig. 5 shows the fields in the bin 6.0-8.8 MeV. The neutron flow
dominates in both interior and surface. Since  protons and
neutrons move in phase, the total flow is essentially isoscalar, see
panels c) and d) in comparison. The T=0 character of the lower bin of
LES is in accordance with previous theoretical results
\cite{Vr01,Paar09} and experimental findings, e.g.  for $^{124}$Sn
\cite{En10}. What is important for our aims, Fig. 5 clearly
demonstrates the overwhelming toroidal flow in neutron and T=0 cases
(less in the proton case). This is in accordance with the TR(T=0)
strength from Fig. 2b, which is strictly peaked just at 7-8 MeV and
dominates over the CR(T=0).  Therefore, LES at 6.0-8.8 MeV is of almost
pure toroidal (vortical) nature! The irrotational
PDR flow is not seen at all.

The LES fields in the bin 8.8-10.5 MeV in Fig. 6 are more complicated. The
flow is mainly isovector in the nuclear interior and isoscalar at the
surface, thus demonstrating an isospin-mixed character (again in
accordance to RMF findings \cite{Vr01,Paar09}).  Furthermore, the TR
flow is faint. Actually there are hints of several flows: TR (b-c), CR
(b), and familiar linear dipole (a,d). This complex picture reflects
the fact that, following Figs. 2b and c, the region 8.8-10.5 MeV hosts
various modes and feels already the vicinity to the GDR.

Finally, Fig. 7 exhibits the $r^2$-weighted CTD to highlight the
role of surface nucleons (e.g. the neutron excess) in the
peripheral reactions like $(\alpha, \alpha')$ and, to a lesser
extent, photo-absorption. Flows in Fig. 7 correspond to the TD in
Fig. 3 (a-b). The $r^2$-weighted presentation weakens the interior
flow and emphasizes the role of the neutron excess, see Fig.7 b).
Nevertheless, the LES still keeps their TR and mixed (TR/CR)
nature in both bins, 6.0-8.8 and 8.8-10.5 MeV. Again we cannot
find any sizable evidence for PDR flow.

The question remains how to observe the velocity fields
experimentally and thus disclose the true nature of LES. The
typical reactions mentioned above are mainly sensitive to the
nuclear surface and lose the important information on the nuclear
interior. This is especially the case for the isoscalar $(\alpha,
\alpha')$ whose response is driven by the operator $r^3 Y_1$ with
a huge surface enhanced factor. (Note also that the most relevant
$(\alpha, \alpha')$ measurements of ISGDR in $^{208}$Pb
\cite{Uchida_PRC_04} consider the energy interval $\omega>$8 MeV
and so, following our calculations,  lose the TR(T=0) peaked at
7-8 MeV).  Perhaps, the $(e,e')$ reaction which can cover both
nuclear surface and interior is the most promising tool to examine
LES flows.

In this study we do not take into account the coupling with complex
configurations which may be essential for LES \cite{En10,Li08}. However
the TR/CR signatures in LES look strong enough to be appreciably
smeared out by this effect.

In conclusion, Skyrme-RPA calculations have been performed to inspect
the nature of the E1 low-energy strength (LES), often denoted as the
pygmy dipole resonance (PDR) and associated with the picture that the
neutron skin oscillates against the nuclear core. Strength functions,
averaged transition densities and averaged current fields (collecting
contributions of all RPA states in a given energy interval) were used
for the analysis. The current fields turned out to be most important
to illustrate the LES flows. The results show that, in agreement with
previous studies \cite{Vr01,Paar09}, LES may be divided into two energy
regions, 6.0-8.8 MeV and 8.8-10.5 MeV in our case, where the lower one
is basically isoscalar and higher one is isospin-mixed.

What is most interesting, LES at 6.0-8.8 MeV shows a clear toroidal
(vortical) nature while the interval 8.8-10.5 MeV gives a mixed
toroidal/compression/linear flow. No convincing indicator of PDR-like
flow is found. This means that the familiar treatment of LES as the
out-of-phase motion of the neutron excess against the nuclear core
(arising from the analogy with light halo nuclei and suggested from
$r^2$-weighted transition densities) is misleading. Our study does not
deny the important contribution of the neutron excess to various
(basically peripheral) reactions. At the same time, we find that LES
flow pattern is far from a simple PDR picture and actually involve
various types of motion, irrotational (compression) and vortical
(toroidal). In particular, LES at 6.0-8.8 MeV constitutes an almost
pure toroidal T=0 resonance. This conclusion may have far-reaching
consequences for further exploration of LES and related observables.

The work was partly supported by the DFG RE-322/12-1, Heisenberg-Landau
(Germany - BLTP JINR), and Votruba - Blokhintsev (Czech Republic - BLTP JINR)
grants. P.-G.R. is grateful for the BMBF support under contracts 06
DD 9052D and 06 ER 9063. The support of the research plan MSM 0021620859
(Ministry of Education of the Czech Republic) is appreciated.

\end{document}